\documentclass[pla,twocolumn,showpacs,showkeywords,preprintnumbers,amsmath,amssymb]{revtex4}
\usepackage{tipa}
\usepackage{amssymb}
\usepackage{txfonts}

\usepackage{hyperref}
\usepackage{graphicx}
\usepackage{epsfig}
\begin{document}

\title{Remote atom entanglement in a fiber-connected three-atom system}
\author{Yan-Qing Guo\footnote{
Corresponding author: yqguo@newmail.dlmu.edu.cn}}
\affiliation{Department of Physics, Dalian Maritime University,
Dalian, Liaoning 116026, P.R.China}
\author{Jing Chen} \affiliation{College of
Science, Beijing Forestry University, Beijing 100083, P.R.China}
\author{He-Shan
Song} \affiliation{Department of Physics, Dalian University of
Technology, Dalian, Liaoning 116023, P.R.China}

\begin{abstract}
An Ising-type atom-atom interaction is obtained in a
fiber-connected three-atom system. The interaction is effective
when $\Delta\approx \gamma _{0}\gg g$. The preparations of remote
two-atom and three-atom entanglement governed by this interaction
are discussed in specific parameters region. The overall two-atom
entanglement is very small because of the existence of the third
atom. However, the three-atom entanglement can reach a maximum
very close to $1$.
\end{abstract}
\maketitle

\section{Introduction}
Generating the entanglement between spatially separated atoms
plays an important role in quantum information processing and
quantum computation, such as quantum storage$^{[1]}$, quantum key
distribution$^{[2]}$ and quantum states swapping$^{[3]}$. To
efficiently entangle two or more distant atoms, one must create
some kind of direct or indirect interaction between them, such as
by adopting appropriate measurement on optical fields that
conditionally interact with atoms and thereby the atoms (as a
subsystem) can be projected to an entangled state, or by using
quantum-correlated fields interacting with atoms and thereby the
entanglement among the fields can be transferred to atoms. Based
upon this, a variety of schemes for entangling distant atoms or
distant photons have been proposed recently$^{[4-14]}$. For
example, fascinating schemes have been presented to efficiently
entangle distant atoms, where the single-photon interference
effect was applied with$^{[4]}$ or without$^{[5]}$ weak driving
laser pulse. Recently, S. Mancini and S. Bose proposed a novel
scheme to directly entangle two atoms trapped in distant
cavities$^{[6]}$ which were connected via optical fibers. Using
input-output theory, under adiabatic approximation, the authors
obtained an effective Ising model for two atoms. In their scheme,
photon acted as an intermediate quantum information carrier and
mapped the quantum information from the atom in one cavity to that
in another. Such systems are meaningful not only in quantum
measurement or testing Bell's inequalities but also in potential
applications such as quantum encryption$^{[15]}$ or constructing
universal quantum gates$^{[16]}$ that are essential for designing
quantum network. Nevertheless, in discussing quantum networking
with trapped atoms and photons in cavity QED system$^{[17]}$, two
problems should be overcome: How to generate the entanglement of a
N-atom system? What is the exact influence of the collective
interaction on the entanglement shared by remote atoms? These
problems have been discussed intensively, for instance in the
scheme proposed by Cabrillo et al$^{[4]}$. The simplest multi-atom
case is a three-atom system which might be an intuitive extension
from a two-atom case. In our scheme, We extend the model of
two-atom circumstance in Ref. [6] to three-atom which turns out to
be a three-atom Ising model. Such an approach might be meaningful
in discussing the above problems for multiple distant atoms. We
firstly investigate the dependence of the effective Ising coupling
coefficients on the atom-cavity detuning and cavity leakage. Then,
we discuss the influence of an atom on the other two atoms
entanglement properties. Furthermore, we study the characters of
remote three-atom entanglement and the tangle between one atom and
the rest two atoms.

\section{Optical fibers connected three-atom system}

The schematic setup for our system is shown in Fig. 1. Three
identical two-level atoms 1, 2 and 3 are trapped in spatially
distant cavities $C_{1}$, $C_{2}$ and $C_{3}$ respectively. All
the cavities are assumed to be single-sided ones. Three
off-resonant external driving field $\varepsilon _{1}$,
$\varepsilon _{2}$ and $\varepsilon _{3}$ are applied upon cavity
$C_{1}$, $C_{2}$ and $C_{3}$ respectively. In each cavity, a local
laser field that is resonantly coupled to the atom is applied. Two
neighboring cavities are connected via optical fibers. Apparently,
the subsystem constituted by cavities $C_{1}$ and $C_{2}$ or
$C_{2}$ and $C_{3}$ is just the setup proposed in Ref. [6].

\begin{figure}
\epsfig{file=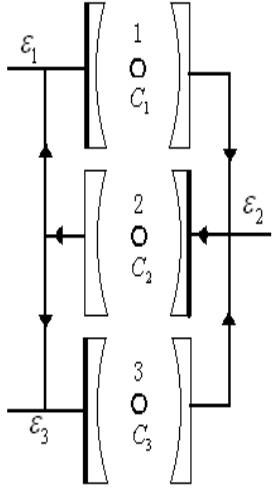, width=4cm, height=7cm, bbllx=45, bblly=19,
bburx=173, bbury=186}\caption{{\protect\footnotesize {Schematic
setup of the system. Three two-level atoms locate in spatially
separated single-sided optical cavities that are connected via
optical fibers.}}}
\end{figure}

In the interaction picture, using cavity input-output
theory$^{[18]}$ and taking adiabatic approximation$^{[19]}$, we
obtain an effective Hamiltonian for this system as (see Appendix
A)
\begin{equation}
H_{eff}=J_{12}\sigma _{1}^{z}\sigma _{2}^{z}+J_{23}\sigma
_{2}^{z}\sigma _{3}^{z}+J_{31}\sigma _{3}^{z}\sigma
_{1}^{z}+\sum\limits_{i}\Gamma _{i}(\sigma _{i}^{+}+\sigma
_{i}^{-}),
\end{equation}
which is a three-particle Ising chain with \textbf{magnetic
fields} perpendicular to the $z$ direction$^{[20]}$, $J_{12}$ and
$J_{23}$ represent the nearest-neighbor (NN) atoms coupling
coefficients, while $J_{31} $ represents next-nearest-neighbor
(NNN) atoms interaction strength. $\sigma _{i}^{z} (i=1,2,3)$ is
spin operator of atom $i$, $\sigma _{i}^{+}(\sigma _{i}^{-})$ is
atomic raising (lowering) operators. $\Gamma _{i}$ is the
magnitude of the locally applied laser field interacting with atom
$i$. We define
\begin{center}
\begin{align}
J_{12}=&2\gamma _{0}\chi ^{2}Im\left\{ \alpha_{1} \alpha_{2}
^{\ast }e^{i\phi_{21}}/[M^{2}-W^{2}]\right\},  \nonumber \\
J_{23}=&2\gamma _{0}\chi ^{2}Im\left\{ \alpha_{3} \alpha_{2}
^{\ast}e^{i\phi_{32}}/[M^{2}-W^{2}]\right\},  \nonumber \\
J_{31}=&2\gamma _{0}\chi ^{2}Im\left\{ \gamma _{0}\alpha_{3}
\alpha_{1} ^{\ast
}e^{i(\phi_{23}+\phi_{12})}/[M(M^{2}-W^{2})]\right\},
\end{align}
\end{center}
where $\gamma _{0}$ is the cavity leakage rate, $\chi =g^{2}/
\Delta $, $g$ is the coupling strength between the atom and the
cavity field in cavity $C_{i}$, $\Delta $ is the detuning between
the atomic internal transition and cavity field frequency, where,
large detuning approximation has been assumed, i.e. $\Delta\gg g$,
and $M=i\Delta +\gamma_{0}$, $W^{2}=\gamma _{0}^{2}\left[
e^{i(\phi _{21}+\phi _{12})}+e^{i(\phi _{32}+\phi _{23})}\right]$.
The phase factors $\phi _{ij}(ij=12,21,23,32)$ are caused from the
photons transmission along optical fibers from cavity $C_{j}$ to
cavity $C_{i}$$^{[21]}$. Physically, they depend on the frequency
of the photons and the distance between cavities. And
\begin{align}
\alpha _{1} &=\frac{\varepsilon _{1}M^{2}+\varepsilon_{2}M\gamma
_{0}e^{i\phi _{12}}+\gamma _{0}^{2}[\varepsilon
_{3}e^{i\Theta_{13}}-\varepsilon
_{1}e^{i\Phi_{1}}]}{M(M^2-W^2)},  \nonumber \\
\alpha _{2} &=\frac{\varepsilon _{2}M+\gamma _{0}(\varepsilon
_{1}e^{i\phi
_{21}}+\varepsilon _{3}e^{i\phi _{23}})}{M^2-W^2},\nonumber \\
\alpha _{3} &=\frac{\varepsilon _{3}M^{2}+\varepsilon_{2}M\gamma
_{0}e^{i\phi _{32}}+\gamma _{0}^{2}[\varepsilon
_{1}e^{i\Theta_{31}}-\varepsilon _{3}e^{i\Phi_{3}}]}{M(M^2-W^2)},
\end{align}
where $\Theta_{13}=\phi _{12}+\phi _{23}, \Phi _{1}=\phi
_{23}+\phi _{32}, \Theta_{31}=\phi _{32}+\phi _{21}, \Phi_{3}=\phi
_{21}+\phi _{12}$. The global system is now determined by a series
of independent parameters as $\varepsilon _{1},\varepsilon _{2},
\varepsilon _{3}, \Delta, \gamma _{0}, \phi _{12}, \phi _{32},
\Gamma _{1}, \Gamma _{2}$ and $\Gamma_{3}$. In next section, we
discuss the optimal region of the parameters for the preparation
of remote atom entanglement.

\section{Parameters space description of Ising coupling coefficients}
From Eqs. (3), the condition $M^2 \approx W^2$ leads to large
Ising coupling coefficients. This condition also keeps the
validity of the adiabatic approximation in case of weak local
laser fields, i.e. $J_{12}(J_{23}, J_{31})\gg \Gamma_{i},
i=1,2,3$. We can further simplify the condition as
\begin{eqnarray}
\phi _{21}+\phi _{12}\approx\phi _{23}+\phi
_{32}\approx\frac{\pi}{2}, \Delta\approx\gamma_{0}.
\end{eqnarray}
For simplicity, we assume $\varepsilon _{1}=\varepsilon
_{2}=\varepsilon _{3}=\varepsilon _{0}$,
$\Gamma_{1}=\Gamma_{2}=\Gamma_{3}=\Gamma_{0}$. The parameters
space can now be expressed in unit of $g$ as $(\Delta/g, \gamma
_{0}/g)g$. In Fig. 2-3, we give the description of Ising coupling
coefficients for NN and NNN atoms in the parameters space. Where
we assume $\varepsilon _{0}=2g$. We can see that the coupling
coefficients for NN atoms as well as NNN atoms can be divided into
two regions: (a) the region where $\Delta>\gamma_{0}$, (b) the
region where $\Delta<\gamma_{0}$. In most area of the two regions,
the coupling coefficients are very small, only in the regions just
besides the line $\Delta=\gamma _{0}$ are they large enough so
that the validity of the adiabatic approximation can be kept.

\begin{figure}
\epsfig{file=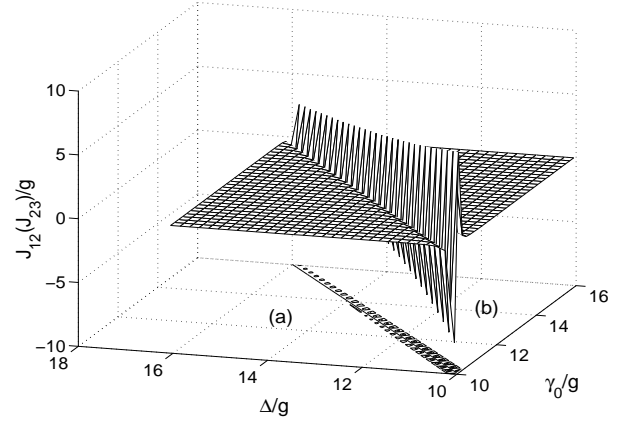, width=8cm,
height=6cm}\caption{{\protect\footnotesize {The amount of NN atoms
coupling coefficients($J_{12}(J_{23})$) under different detuning
($\Delta$) and cavity leakage rate ($\gamma_{0}$). All the
variables are presented in unit of $g$.}}}
\end{figure}

\begin{figure}
\epsfig{file=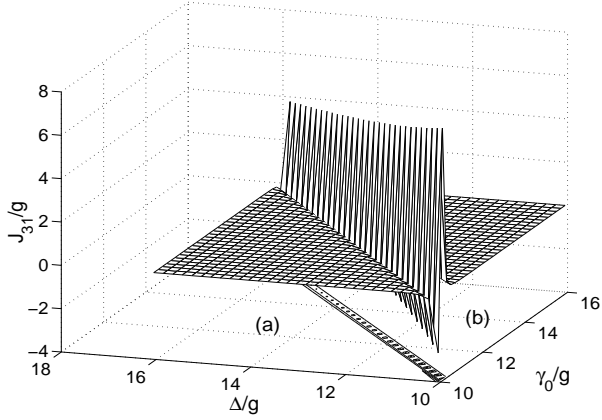, width=8cm,
height=6cm}\caption{{\protect\footnotesize {The amount of NNN
atoms coupling coefficients($J_{31})$ under different detuning
($\Delta$) and cavity leakage rate ($\gamma_{0}$). All the
variables are presented in unit of $g$.}}}
\end{figure}

In the following discussions, we will study two-atom entanglement
nature and three-atom entanglement properties based on the
parameter space.
\section{Nearest-neighbor and next-nearest-neighbor remote two-atom entanglement}
In this section, we discuss the nature of remote two-atom
subsystem entanglement which is generated in our system. Wootters
proposed a general measurement for the amount of two-qubit (noted
as 1 and 2) entanglement. It is named as Concurrence$^{[22]}$:
\begin{equation}
C_{12}=\max \{0,\lambda _{1}-\lambda _{2}-\lambda _{3}-\lambda
_{4}\},
\end{equation}%
where $\lambda _{i}$ are the non-negative square roots of the four
eigenvalues of non-Hermitian matrix $\rho_{12} \tilde{\rho}_{12}$
with $\tilde{\rho}_{12}$ defined as $(\sigma _{y}\otimes \sigma
_{y})\rho^{\ast }_{12} (\sigma _{y}\otimes \sigma _{y})$, where
$\rho_{12}$ is the density matrix of the two-qubit system.

We depict the two-atom entanglement situation in Fig. 4-5 for
different parameter spaces. We assume that all the atoms are
initially in their ground state, so that $|\psi
(0)\rangle=|g\rangle _{1}\otimes |g\rangle _{2}\otimes |g\rangle
_{3}$.

\begin{figure}
\epsfig{file=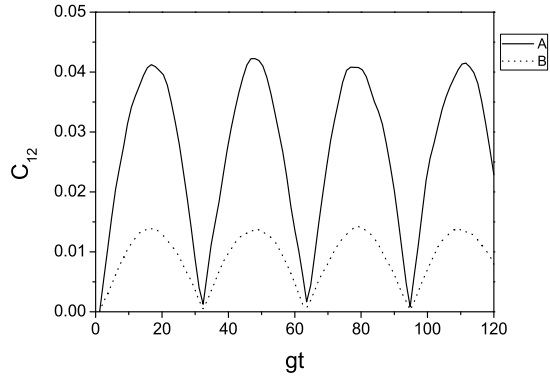, width=8cm,
height=6cm}\caption{{\protect\footnotesize {Entanglement of atom 1
and 2 versus time for A: (solid line) $J_{31}\approx1.2g$, B:
(dotted line) $J_{31}\approx-1.2g$. Where we assume
$\Gamma_{0}=0.1g$.}}}
\end{figure}

\begin{figure}
\epsfig{file=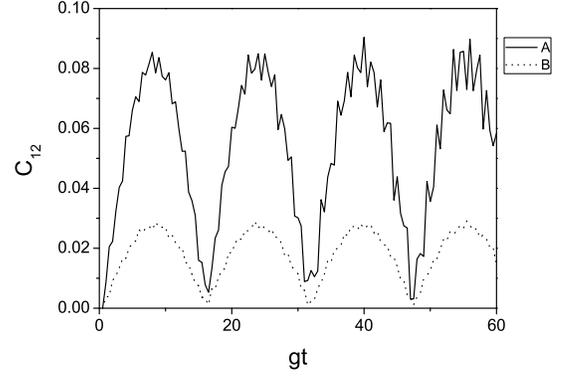, width=8cm,
height=6cm}\caption{{\protect\footnotesize {Descriptions are same
as those in Fig. 4 but for $\Gamma_{0}=0.2g$. Note that the whole
time scale is half of that in Fig. 4.}}}
\end{figure}

Firstly, we investigate the influence of $J_{31}$ on the
entanglement of NN atoms. In Fig. 4, we adopt appropriate values
of $\Delta$ and $\gamma _{0}$ (which satisfy the condition in Eq.
(4)) and assume $\Gamma_{0}=0.1g$. The Ising coupling coefficients
are $J_{12}=J_{23}\approx-2.4g$, $J_{31}\approx1.2g$ for solid
line, and $J_{31}\approx-1.2g$ for dotted line. If all the signs
of the coefficients are reversed, the resulting concurrences are
not changed. Evidently, relative larger entanglement for NN atoms
can be obtained when $J_{12}(J_{23})\cdot J_{31}<0$. While,
compared with the result in Ref. [6], the overall entanglement is
very weak since two-atom subsystem is in mixed state during the
evolution.

In addition, the NN atoms entanglement can be manipulated through
the alternating of the locally applied laser fields. In Fig. 5, we
adopt the same parameters as those in Fig. 4 but for
$\Gamma_{0}=0.2g$. Fig. 5 indicates that, the increase of
$\Gamma_{0}$ remarkably improves the NN atoms entanglement. The
period is depressed, but the amount of entanglement is much
enhanced. The amount of entanglement for NNN atoms, under the
parameters we assumed, is generally much weaker than NN atoms. To
improve the entanglement for NNN atoms, The Ising coupling
coefficient between NNN atoms must be enhanced. In Fig. 6, we
depict the entanglement for NNN atoms. Correspondingly,
$J_{12}=J_{23}=-2.4g$, $J_{31}=1.2g$. To modulate the
entanglement, we let $\Gamma _{2}=0$. Under this circumstance, the
entanglement for NNN atoms can compare with that for NN atoms (see
Fig. 4).
\begin{figure}
\epsfig{file=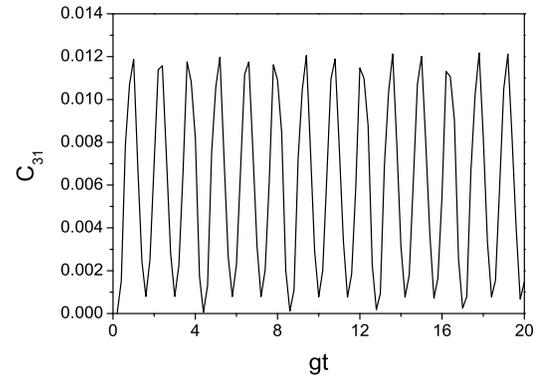, width=8cm,
height=6cm}\caption{{\protect\footnotesize {Entanglement of atom 1
and 3 versus time when the laser field applied in cavity $C_{2}$
is turn off, that is $\Gamma_{1}=\Gamma_{3}=0.3g$,
$\Gamma_{2}=0$.}}}
\end{figure}

\section{The remote three-atom entanglement properties}
The intrinsic three-partite entanglement which is widely used for
measuring three-partite entanglement of pure states is defined
as$^{[23]}$
\begin{equation}
C_{123}=C_{1(23)}-C_{12}^{2}-C_{13}^{2},
\end{equation}
where $C_{1(23)}$, which represents the tangle between a subsystem
1 and the rest of the global system (denoted as $(23)$), is
written as
\begin{equation}
C_{1(23)}=4Det\rho_{1}=2(1-Tr\rho _{1}^{2}).
\end{equation}
In Fig. 7, we plot the remote three-atom entanglement
$C_{123}$(the solid line), the tangle $C_{1(23)}$(the dotted
line), and the Concurrence $C_{12}$(the dashed line) for $|\psi
(0)\rangle=|g\rangle _{1}\otimes |g\rangle _{2}\otimes |g\rangle
_{3}$. Where the corresponding parameters are same as those of the
dotted line in Fig. 5. In fact, there is only very little
difference between $C_{123}$ and $C_{1(23)}$. To distinguish
$C_{1(23)}$ from $C_{123}$, the line of $C_{1(23)}$ is raised to
$0.1+C_{1(23)}$.

\begin{figure}
\epsfig{file=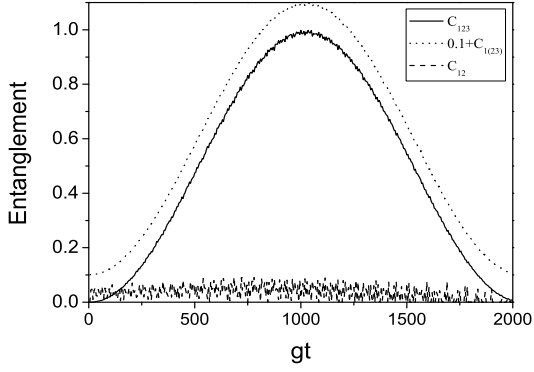, width=8cm,
height=6cm}\caption{{\protect\footnotesize {The three-atom
entanglement $C_{123}$(solid line), the tangle $C_{1(23)}$(dotted
line), and the Concurrence $C_{12}$(dashed line) versus $gt$,
where $J_{12}(J_{23})=-2.4g$, $J_{31}=-1.2g$ and
$\Gamma_{0}=0.2g$. To distinguish $C_{1(23)}$ from $C_{123}$, the
line for $C_{1(23)}$ is raised to $0.1+C_{1(23)}$.}}}
\end{figure}

It has been pointed in last section that the Concurrence that
represents the bipartite entanglement between atom 1 and atom 2 is
very small. While, the tangle, which expresses the entanglement of
atom 1 and the rest of the global system, and the remote
three-atom entanglement can reach maximum values almost 1 for
intermediate values of $gt$. Under the condition of strong Ising
coupling coefficients $(J_{12}(J_{23}, J_{31})\gg\Gamma _{0})$, if
we express the tangle between atom 1 and the rest of the global
system as a sum of the remote three-atom entanglement and the
bipartite entanglement, the three-atom entanglement will act as
the largest contribution. It has been concluded that in a system
of $N$ spin-half particles, under the condition of strong Ising
coupling coefficients, the $N$-partite entanglement will be
dominant$^{[24]}$.
\section{Conclusion}
We have obtained a three-atom Ising chain in cavity QED system by
connecting three distant cavities via optical fibers. The Ising
coupling coefficients are found to be large in the region where
$\Delta\approx \gamma _{0}\gg g$, which keeps the validity of the
adiabatic approximation. We have discussed the generation of
remote atom entanglement. The overall two-atom entanglement is
very small because of the existence of the third atom. While, the
NN atoms entanglement can be improved when the coupling
coefficient of NNN atoms has a contrary sign with respect to that
of NN atoms. The locally applied laser fields play an important
role in modulating the entanglement quantitatively and
qualitatively not only for NN atoms but also for NNN atoms.
Furthermore, we have studied the remote three-atom entanglement
and the tangle. It is shown that three-atom entanglement, which
has a much longer period than two-atom entanglement, can reach a
maximum very close to $1$.

In addition, it should be noted that the dissipation of the photon
information along the fibers should be investigated, while, the
dissipation can be included in the Ising coupling coefficients and
act as a decaying exponential factor $e^{-\nu L}$, where $\nu$ is
the dissipation rate per meter, $L$ is the total length of the
fiber$^{[25]}$. The phase factors $e^{i\phi_{12}}$ and
$e^{i\phi_{23}}$ are then replaced by $e^{i\phi_{12}-\nu L_{12}}$
and $e^{i\phi_{23}-\nu L_{23}}$. In fact, the dissipative effect
along fibers can be compensated by lowering the detuning $\Delta$.
One can obtain large Ising coupling coefficients by adopting the
parameters in the regions just besides the line $\Delta \approx
\sqrt{2e^{-\nu(L_{12}+L_{23})}-1}\gamma _{0}$.
\section*{acknowledgments}
This work was supported by the National Natural Science Foundation
of China under Grant Nos. 10647107 and 10575017.
\section*{Appendix A}
In the interaction picture, the Hamiltonian of the global system
can be written as
\begin{displaymath}
H_{int}=H_{int1}+H_{int2}+H_{int3}+H_{int4},\tag{A1}
\end{displaymath}
where $H_{int1}$ represents the effective interaction of atoms and
cavity fields, $H_{int2}$ is the coupling between external driving
fields and cavity fields , $H_{int3}$ represents the interaction
of locally applied laser fields and atoms, $H_{int4}$ is the
interaction of cavity fields and their environment which is
described as a superposition of series of harmonic oscillators.
Under the condition of large detuning, we have$^{[26]}$
\begin{displaymath}
H_{int1}=\chi\sum\limits_{i}A_{i}^{+}A_{i}\sigma _{i}^{z},\tag{A2}
\end{displaymath}
where i=1,2,3, $A_{i}(A_{i}^{+})$ represent cavity fields
annihilation (creation) operators in cavities $C_{i}$.
And$^{[18]}$
\begin{align}
H_{int2}=\sum\limits_{i}\varepsilon _{i}(A_{i}^{+}+A_{i}),\tag{A3}
\end{align}
\begin{align}
H_{int3}=\sum\limits_{i}\Gamma _{i} (\sigma _{i}^{+}+\sigma
_{i}^{-}).\tag{A4}
\end{align}

We assume $\Gamma_{i}$ are weak enough so that the quantum
adiabatic theory$^{[19]}$ can be applied in the following
calculations.
\begin{align}
H_{int4} =i\int\limits_{-\infty }^{+\infty }d\omega \
{\sum\limits_{i}{\kappa_{C_{i}}[ b_{C_{i}}(\omega
)A_{i}^{+}+h.c.]}},\tag{A5}
\end{align}
where $b_{C_{i}}(\omega)$, $i=1,2,3$, are the annihilation
operators of the harmonic oscillators with frequency $\omega$.
$\kappa_{C_{i}}$ are the interaction strengths between cavity
$C_{i}$ and the harmonic oscillators. The kinetic equations for
cavity field operators turn out to be$^{[18]}$
\begin{align}
\dot{A_{1}} &=-(i\Delta+i\chi\sigma _{1}^{z}+\frac{\gamma
_{C_{1}}}{2})A_{1}+\sqrt{
\gamma _{C_{1}}}A_{1,in}+\varepsilon _{1},\nonumber\\
\dot{A_{2}} &=-(i\Delta+i\chi\sigma _{2}^{z}+\frac{\gamma
_{C_{1}}}{2})A_{2}+\sqrt{\gamma
_{C_{2}}}A_{2,in}+\varepsilon _{2},   \nonumber\\
\dot{A_{3}} &=-(i\Delta+i\chi\sigma _{3}^{z}+\frac{\gamma
_{C_{1}}}{2})A_{3}+\sqrt{\gamma _{C_{3}}}A_{3,in}+\varepsilon
_{3},\tag{A6}
\end{align}
where $\gamma _{C_{i}}=2\pi [k_{C_{i}}(\omega )]^{2}$ ($i=1,2,3$).
If cavities $C_{1}$ and $C_{2}$ are connected via optical fibers
(as shown in Fig. 1), so are cavities $C_{2}$ and $C_{3}$, the
input-output conditions should be included, so that$^{[21]}$
\begin{align}
\dot{A_{1}} &=-\frac{\gamma _{C_{1}}}{2}A_{1}+\sqrt{\gamma
_{C_{1}}}A_{2,out}e^{i\phi _{12}}
,  \nonumber \\
\dot{A_{2}} &=-\frac{\gamma _{C_{2}}}{2}A_{2}+\sqrt{\gamma
_{C_{2}}}A_{1,out}e^{i\phi _{21}}\nonumber \\
&+\sqrt{\gamma _{C_{2}}}A_{3,out}e^{i\phi
_{23}},  \nonumber \\
\dot{A_{3}} &=-\frac{\gamma _{C_{3}}}{2}A_{3}+\sqrt{\gamma
_{C_{3}}}A_{2,out}e^{i\phi _{32}}.\tag{A7}
\end{align}

For simplicity, assuming the decay rates $\gamma _{C_{1}}=\gamma
_{C_{2}}=\gamma _{C_{3}}=\gamma _{0}$ and taking into account the
usual boundary conditions$^{[18]}$
\begin{align}
A_{i,out}+A_{i,in}=\sqrt{\gamma _{0}}A_{i},\tag{A8}
\end{align}
where $i=1,2,3$, we can rewrite the kinetic equations for cavity
field operators as
\begin{align}
\dot{A_{1}}&=-MA_{1}-i\chi A_{1}\sigma _{1}^{z}+\sqrt{\gamma _{0}}
A_{1,in}\nonumber \\
&+e^{i\phi _{12}}(\gamma _{0}A_{2}-\sqrt{\gamma _{0}}
A_{2,in})+\varepsilon _{1},\nonumber\\
\dot{A_{2}}&=-MA_{2}-i\chi A_{2}\sigma _{2}^{z}+\sqrt{\gamma
_{0}}A_{2,in}\nonumber \\
&+\sum\limits_{j=1,3}{e^{i\phi _{2j}}(\gamma
_{0}A_{j}-\sqrt{\gamma_{0}}A_{j,in})}+\varepsilon _{2}
,\nonumber\\
\dot{A_{3}}&=-MA_{3}-i\chi A_{3}\sigma _{3}^{z}+\sqrt{\gamma
_{0}}A_{3,in}\nonumber \\
&+e^{i\phi _{32}}(\gamma
_{0}A_{2}-\sqrt{\gamma_{0}}A_{2,in})+\varepsilon _{3}.\tag{A9}
\end{align}

To solve these equations explicitly, we firstly obtain the
expectation values of cavity field operators through
\begin{align}
\frac{d\left\langle A_{1}\right\rangle }{dt}=\frac{d\left\langle
A_{2}\right\rangle }{dt}=\frac{d\left\langle A_{3}\right\rangle
}{dt}=0.\tag{A10}
\end{align}

The steady states for cavity fields in $C_{1}$, $C_{2}$ and
$C_{3}$ can be obtained as
\begin{align}
\alpha _{1} &=\frac{\varepsilon _{1}M^{2}+\varepsilon_{2}M\gamma
_{0}e^{i\phi _{12}}+\gamma _{0}^{2}[\varepsilon
_{3}e^{i\Theta_{13}}-\varepsilon
_{1}e^{i\Phi_{1}}]}{M(M^2-W^2)},  \nonumber \\
\alpha _{2} &=\frac{\varepsilon _{2}M+\gamma _{0}(\varepsilon
_{1}e^{i\phi
_{21}}+\varepsilon _{3}e^{i\phi _{23}})}{M^2-W^2},\nonumber \\
\alpha _{3} &=\frac{\varepsilon _{3}M^{2}+\varepsilon_{2}M\gamma
_{0}e^{i\phi _{32}}+\gamma _{0}^{2}[\varepsilon
_{1}e^{i\Theta_{31}}-\varepsilon
_{3}e^{i\Phi_{3}}]}{M(M^2-W^2)},\tag{A11}
\end{align}

where $\Theta_{13}=\phi _{12}+\phi _{23}, \Phi _{1}=\phi
_{23}+\phi _{32}, \Theta_{31}=\phi _{32}+\phi _{21}, \Phi_{3}=\phi
_{21}+\phi _{12}$.

Then, in the regime of strong cavity leakage and large detuning
(which lead to $\gamma_{0}, \Delta>>\chi$), the kinetic Eqs. (A9)
are reformed as the following homogeneous linear equations:
\begin{align}
\dot{a_{1}} &=-Ma_{1}-i\chi \alpha _{1} \sigma _{1}^{z}
+\sqrt{\gamma _{0}}a_{1,in}\nonumber \\
&+e^{i\phi _{12}}(\gamma
_{0}a_{2}-\sqrt{\gamma _{0}}a_{2,in}),  \nonumber \\
\dot{a_{2}} &=-Ma_{2}-i\chi \alpha _{2} \sigma
_{2}^{z}+\sqrt{\gamma _{0}}a_{2,in}\nonumber \\
&+\sum\limits_{j=1,3}{e^{i\phi _{2j}}(\gamma
_{0}a_{j}-\sqrt{\gamma _{0}}
a_{j,in})},\nonumber \\
\dot{a_{3}} &=-Ma_{3}-i\chi \alpha _{3} \sigma _{3}^{z}+\sqrt{
\gamma _{0}}a_{3,in}\nonumber \\
&+e^{i\phi _{32}}(\gamma _{0}a_{2}-\sqrt{\gamma
_{0}}a_{2,in}),\tag{A12}
\end{align}
where we have replaced field operators $A_{i}$ with $a_{i}+\alpha
_{i}$ (i=1,2,3). In solving Eqs. (A12), one can adiabatically
eliminate the effect of vacuum input noise. The resulting cavity
field operators are now represented by linear combinations of
atomic spin operators $\sigma _{i}^{z}$ (i=1,2,3). Substituting
the resulting field operators into Eq. (A1), we get the effective
Hamiltonian of the global system in the interaction picture as
\begin{align}
H_{eff}&=J_{12}\sigma _{1}^{z}\sigma _{2}^{z}+J_{23}\sigma
_{2}^{z}\sigma _{3}^{z}+J_{31}\sigma _{3}^{z}\sigma
_{1}^{z}\nonumber \\
&+\sum\limits_{i}\Gamma _{i}(\sigma _{i}^{+}+\sigma
_{i}^{-}),\tag{A13}
\end{align}
where $J_{12}, J_{23}$ and $J_{31}$ are expressed by Eqs. (2).

In deriving Eq. (A13), we neglect self-energy terms including
$\sigma _{i}^{z}$ and self-interaction terms including $( \sigma
_{i}^{z})^{2}$ and $(\sigma _{i}^{z})^{3}$ that do not change the
initial system state. Also, we eliminate higher order terms that
include $\chi ^{3}\sigma _{1}^{z}\sigma _{2}^{z}\sigma _{3}^{z}$
since the corresponding coupling coefficients are much weaker than
$J_{12}$, $J_{23}$ and $J_{31}$. The typical difference between
this Hamiltonian and that in Ref. [6] lies in the third term in
Eq. (A13).


\begin{thebibliography}{99}

\bibitem{1} van der Wal C H, Eisaman M D, Andre A, Walsworth R L, Phillips D F, Zibrov A S and Lukin M D 2003 \textit{Science} \textbf{301} 196

\bibitem{2} Inoue K, Waks E and Yamamoto Y 2002 \textit{Phys. Rev. Lett.} \textbf{89}
 037902

\bibitem{3} Kuzmich A and Polzik E S 2002 \textit{Phys. Rev. Lett.} \textbf{85}
 5639

\bibitem{4} Cabrillo C, Cirac J, Garcia-Fernandez P and Zoller P
1999 \textit{Phys. Rev. A} \textbf{59} 1025

\bibitem{5} Feng X L, Zhang Z M, Li X D, Gong S Q and Xu Z Z 2003
\textit{Phys. Rev. Lett.} \textbf{90} 217902


\bibitem{6} Mancini S and Bose S 2004 \textit{Phys. Rev. A} \textbf{70} 022307

\bibitem{7} Li H C, Li X H, Lin X, Lin X M and Yang R C 2007 \textit{Chin. Phys.} \textbf{16}
1209

\bibitem{8} Fang M F and Tan J 2006 \textit{Chin. Phys.} \textbf{15} 2514

\bibitem{9} Chimczak G 2005 \textit{Phys. Rev. A} \textbf{71} 052305

\bibitem{10} Duan L M and Kimble H J 2003 \textit{Phys. Rev. Lett.} \textbf{90}
 253601

\bibitem{11} Guo Y Q, Chen J and Song H S 2006 \textit{Chin. Phys.
Lett.} \textbf{23} 1088

\bibitem{12} SimonC and Irvine W T M 2003 \textit{Phys. Rev. Lett.} \textbf{91}
 110405


\bibitem{13} Zou X B, Pahlke K and Mathis W 2003 \textit{Phys. Rev. A} \textbf{68}
 024302

\bibitem{14} Ficek Z and Tana\'{s} R 2003 \textit{quant-ph} 0302124

\bibitem{15} Ekert A 1991 \textit{Phys. Rev. Lett.} \textbf{67} 661

\bibitem{16} Zou X B and Mathis W 2005 \textit{Phys. Rev. A} \textbf{71}
 042334

\bibitem{17} Moehring D L, Madsen M J, Younge K C,
Kohn R N, Jr P Maunz, Duan L M, Monroe C and Blinov B B 2007
\textit{J. Opt. Soc. Am. B} \textbf{24} 300

\bibitem{18} Walls D F and Milburn G J 1994 Quantum Optics (Springer:
Berlin) p121

\bibitem{19} Sanrady M S, Wu L A and Lidar D A 2004 \textit{Quantum
Information Processing} \textbf{3} 331

\bibitem{20} Gunlycke D, Kendon V M and Vedral V 2001 \textit{Phys. Rev. A} \textbf{
64} 042302

\bibitem{21} Wiseman H M and Milburn G J 1994 \textit{Phys. Rev. A} \textbf{49}
 4110

\bibitem{22} Wootters W K 1998 \textit{Phys. Rev. Lett.} \textbf{80} 2245

\bibitem{23} Coffman V, Kundu J and Wootters W K 2000 \textit{Phys. Rev.
A} \textbf{61} 052306

\bibitem{24} \v{S}telmachovi\v{c} P and Bu\v{z}ek V 2004 \textit{Phys. Rev. A}
\textbf{70} 032313

\bibitem{25} Tittel W, Brendel J, Gisin B, Herzog T, Zbinden H and Gisin N 1998 \textit{Phys. Rve. A} \textbf{57}
 3229

\bibitem{26} Holland M J, Walls D F and Zoller P 1991 \textit{Phys. Rev. Lett.}
\textbf{67} 1716


\end{thebibliography}
\end{document}